\documentclass[prl,twocolumn,msmath,superscriptaddress,amssymb,showpacs]{revtex4-1}

\usepackage{color} 
\usepackage{graphicx}
\usepackage{epstopdf}
\usepackage{bm}
\pdfoutput=1
\begin{document}

\DeclareGraphicsExtensions{.eps, .png, .jpg}
\bibliographystyle{prsty}

\title {Electronic correlation assisted ferroelectric metallic state in LiOsO$_3$}

\author{I. Lo Vecchio}
\affiliation{Dipartimento di Fisica, Universit\`{a} di Roma ``Sapienza", Piazzale A. Moro 2, I-00185 Roma, Italy}

\author{G. Giovannetti}
\affiliation{International School for Advanced Studies (SISSA) and CNR-IOM-Democritos National Simulation Centre, Via Bonomea 265, I-34136 Trieste, Italy}
\affiliation{Institute for Theoretical Solid State Physics, IFW-Dresden, PF 270116, 01171 Dresden, Germany}

\author{M. Autore}
\affiliation{INFN and Dipartimento di Fisica, Universit\`{a} di Roma ``Sapienza", Piazzale A. Moro 2, I-00185 Roma, Italy}

\author{P. Di Pietro}
\affiliation{INSTM Udr Trieste-ST and Elettra Sincrotrone Trieste S.C.p.A., Area Science Park, I-34149 Basovizza, Trieste, Italy}

\author{A. Perucchi}
\affiliation{INSTM Udr Trieste-ST and Elettra Sincrotrone Trieste S.C.p.A., Area Science Park, I-34149 Basovizza, Trieste, Italy}

\author{Jianfeng He}
\affiliation{Superconducting Properties Unit, National Institute for Materials Science, 1-1 Namiki, Tsukuba, Ibaraki 305-0044, and Graduate School of Chemical Sciences and Engineering, Hokkaido University, North 10 West 8, Kita-ku, Sapporo, Japan}

\author{K. Yamaura}
\affiliation{Superconducting Properties Unit, National Institute for Materials Science, 1-1 Namiki, Tsukuba, Ibaraki 305-0044, and Graduate School of Chemical Sciences and Engineering, Hokkaido University, North 10 West 8, Kita-ku, Sapporo, Japan}

\author{M. Capone}
\affiliation{International School for Advanced Studies (SISSA) and CNR-IOM-Democritos National Simulation Centre, Via Bonomea 265, I-34136 Trieste, Italy}

\author{S. Lupi}
\affiliation{CNR-IOM and Dipartimento di Fisica, Universit\`a di Roma ``Sapienza", Piazzale A. Moro 2, I-00185, Roma, Italy}
\date{\today}

\begin{abstract}
LiOsO$_3$ has been recently identified as the first unambiguous ``ferroelectric metal", experimentally realizing a prediction from 1965 by Anderson and Blount. In this work, we investigate the metallic state in LiOsO$_3$ by means of infrared spectroscopy supplemented by Density Functional Theory and Dynamical Mean Field Theory calculations. Our measurements and theoretical calculations clearly show that LiOsO$_3$ is a very bad metal with a small quasiparticle weight, close to a Mott-Hubbard localization transition. The agreement between experiments and theory allows us to ascribe all the relevant features in the optical conductivity to strong electron-electron correlations within the $t_{2g}$ manifold of the osmium atoms.
\end{abstract}
\pacs{74.25.71.27.+a,Gz,78.30.-j, 77.80.B-}
\maketitle

\emph{Introduction -} Ferroelectric materials display a spontaneous polarization due to an inversion symmetry breaking induced by the macroscopic ordering of local dipole moments. Multiferroic materials are defined by coexistent and coupled magnetic and ferroelectric order.
It is a natural expectation that  ferroelectric (and consequently multiferroic) ordering can only happen in insulators in order to avoid the metallic charge carriers to screen out the ferroelectric polarization. 
The existence of ``ferroelectric metals" challenging this expectation has been hypothesized in 1965 by Anderson and Blount \cite{Anderson}, who have shown that ferroelectricity can occur in a metal as long as the electrons at the Fermi level are decoupled from the ferroelectric distortions. Under these circumstances, the ferroelectric ordering can take place through a second-order transition.
In 2013 the first unambiguous realization of this proposal has been reported in LiOsO$_3$, where a second-order transition leads to a ferroelectric ionic structure below 140 K while the material remains conducting\cite{Yamaura-13}.
This behavior is accompanied by a large residual resistivity which exceeds by two orders of magnitude that of prototypical metals as gold, and by a Curie-Weiss like behavior of the magnetic susceptibility in the ordered phase which suggests the presence of almost localized magnetic moments, characteristic precursors of Mott localization\cite{Yamaura-13}.

First theoretical insights about the mechanism behind the ferroelectric instability and the metallic character of LiOsO$_3$ were gained by using Density Functional Theory (DFT) calculations \cite{Kim-14,Giovannetti-14,Liu-15}.
The ferroelectric transition in metallic LiOsO$_3$ mainly results from the cooperative displacements of Li and O ions, while the metallic conduction is associated with the  $t_{2g}$ orbitals of osmium which are partially occupied by three electrons per ion\cite{Giovannetti-14}. In this half-filling configuration electron-electron interactions are particularly effective and can lead to a highly correlated metal with poor metallic properties and incipient local magnetic moments.

Although the coexistence of metallic conduction and ferroelectric order in LiOsO$_3$ can be understood in terms of the longstanding proposal by Anderson and Blount \cite{Anderson}, which relies on weak coupling between the electrons at the Fermi level and the soft phonon responsible for removing inversion symmetry \cite{Puggioni}, the fingerprints of strong correlations and their connection with the ferroelectric transition deserve further investigation and understanding. Indeed strong correlations are relevant also in previously known non-centrosymmetric metals. For instance LaSr$_2$Cu$_2$GaO$_7$ shares a very similar structure and the $d^9$ electronic configuration for Cu atoms with the high-temperature superconductor YBa$_2$Cu$_3$O$_7$ in which strong correlation effects are unambiguously recognized \cite{Poeppelmeier-91}. In the pyrochlore oxide Cd$_2$Re$_2$O$_7$ a piezoelectric transition with lifting of the inversion symmetry takes place\cite{Sergienko-04} and a strong mass renormalization is measured\cite{Wang-02}.

In this work we elucidate the strongly correlated nature of the ferroelectric metallic state in LiOsO$_3$ by means of optical conductivity measurements and theoretical calculations based on the merger of DFT\cite{DFT} and Dynamical Mean Field Theory (DMFT)\cite{DMFT}.
Theory and experiments provide striking evidence that LiOsO$_3$ is a strongly correlated metal, thereby strengthening the link between the bad metallic behavior brought by short-range Coulomb interaction and ferroelectric metals. We argue that strong correlations, leading to a bad metallic behavior, cooperate with the Anderson-Blount mechanism. The bad metal is indeed much less effective than a conventional one in screening the electric dipoles, and it can effectively decouple the nearly localized electrons at the Fermi level from the distortions responsible for ferroelectricity.

%\emph{Sample growth -} High density polycrystalline pellets of LiOsO$_3$ were prepared by solid-state reaction under high pressure \cite{Yamaura-13}. Trivial degree of Li off-stoichiometry was suggested in a former study by a chemical method; however, it was not confirmed by neutron diffraction \cite{Yamaura-13}. \textcolor{red}{Sample dependence was indeed trivial on the lattice and electronic properties}, indicating that the impact of possible Li off-stoichiometry on the bulk characterization was insignificant. Therefore, we have used the stoichiometric composition (LiOsO$_3$) throughout the experimental analysis and calculations.
\emph{Sample growth -} High-density polycrystalline pellets of LiOsO$_3$ were prepared by solid-state reaction under high-pressure \cite{Yamaura-13}. A small Li deficiency (Li$_{0.98}$OsO$_3$) was suggested in a former study by a chemical method, however, it was not confirmed by further experiments \cite{Yamaura-13}. This indicates that the impact of possible Li off-stoichiometry on the bulk properties is insignificant. Therefore, we have used the stoichiometric composition (LiOsO$_3$) throughout the experimental analysis and calculations.

\begin{figure}[t]
\begin{center}
\leavevmode
\includegraphics [width=7cm,angle=-90]{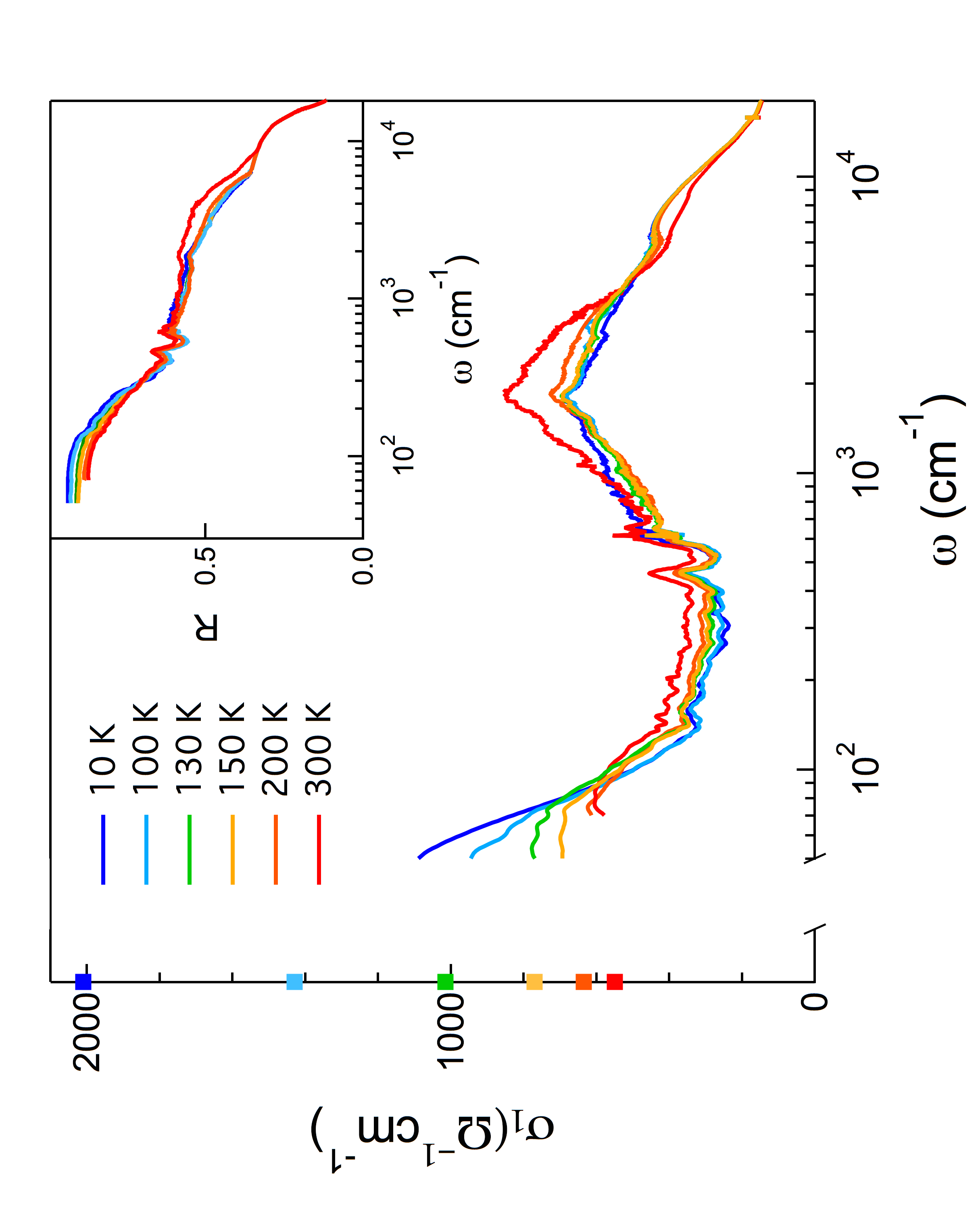}  
\end{center}
\caption{Optical conductivity of LiOsO$_3$ from 10 K to room temperature on a log frequency scale. The structural ferroelectric transition temperature is 140 K. Symbols on the left axis represent dc values measured on a sample from the same batch. The near-normal incidence reflectance plotted in the inset was used to perform Kramers-Kronig transformations.}
\label{Fig1}
\end{figure}

\emph{Optical Measurements -} In the inset of Fig. \ref{Fig1} we show the near-normal incidence reflectance of LiOsO$_3$ measured at the SISSI beamline of Elettra synchrotron from 10 K to room temperature \cite{SISSI}. The surface of a high-density polycrystalline pellet was accurately polished and a gold (or silver) surface was evaporated $in$ $situ$ over the sample and used as a reference. A Michelson interferometer was used in the frequency range \cite{note70} from 50 cm$^{-1}$ to 18000 cm$^{-1}$.  The low temperature reflectance R($\omega$) shows a metallic response, approaching unity at zero frequency. Raising the temperature causes an increase of the resistivity and thus a depletion of the metallic behavior. All the curves merge together in the visible frequency range.

In order to obtain the optical conductivity, we performed Kramers-Kronig transformations. Low-frequency reflectance data were extrapolated with Hagen-Rubens method taking into account resistivity dc values measured in a sample coming from the same batch. A $\omega^{-4}$ high-frequency tail was instead merged to the data above 18000 cm$^{-1}$. The result is shown in the main panel of Fig. \ref{Fig1} on a logarithmic scale. 
The symbols on the vertical axis indicate the dc conductivity values calculated from resistivity data \cite{Yamaura-13}. At all temperatures there is a good agreement between the zero frequency limit of $\sigma_1(\omega)$ and the measured $\sigma_{dc}$.
The optical conductivity at 10 K shows a very narrow Drude peak with a minimum around 150 cm$^{-1}$. A phononic region with approximately ten infrared active modes is well noticeable in the 150-700 cm$^{-1}$ frequency range \cite{Kim-14}. By heating the sample the metallic contribution decreases in the far-infrared through a transfer of spectral weight (SW) to a mid-infrared (MIR) band centered around 1800 cm$^{-1}$. It is then interesting to observe that the structural phase transition at 140 K is accompanied by a rather smooth crossover from a coherent Fermi liquid at low temperature to a ``bad metal" at high temperature with an almost depleted Drude contribution.
The total SW is recovered above 10000 cm$^{-1}$. 

\begin{figure}[t]
\begin{center}
\leavevmode
\includegraphics [width=1\columnwidth]{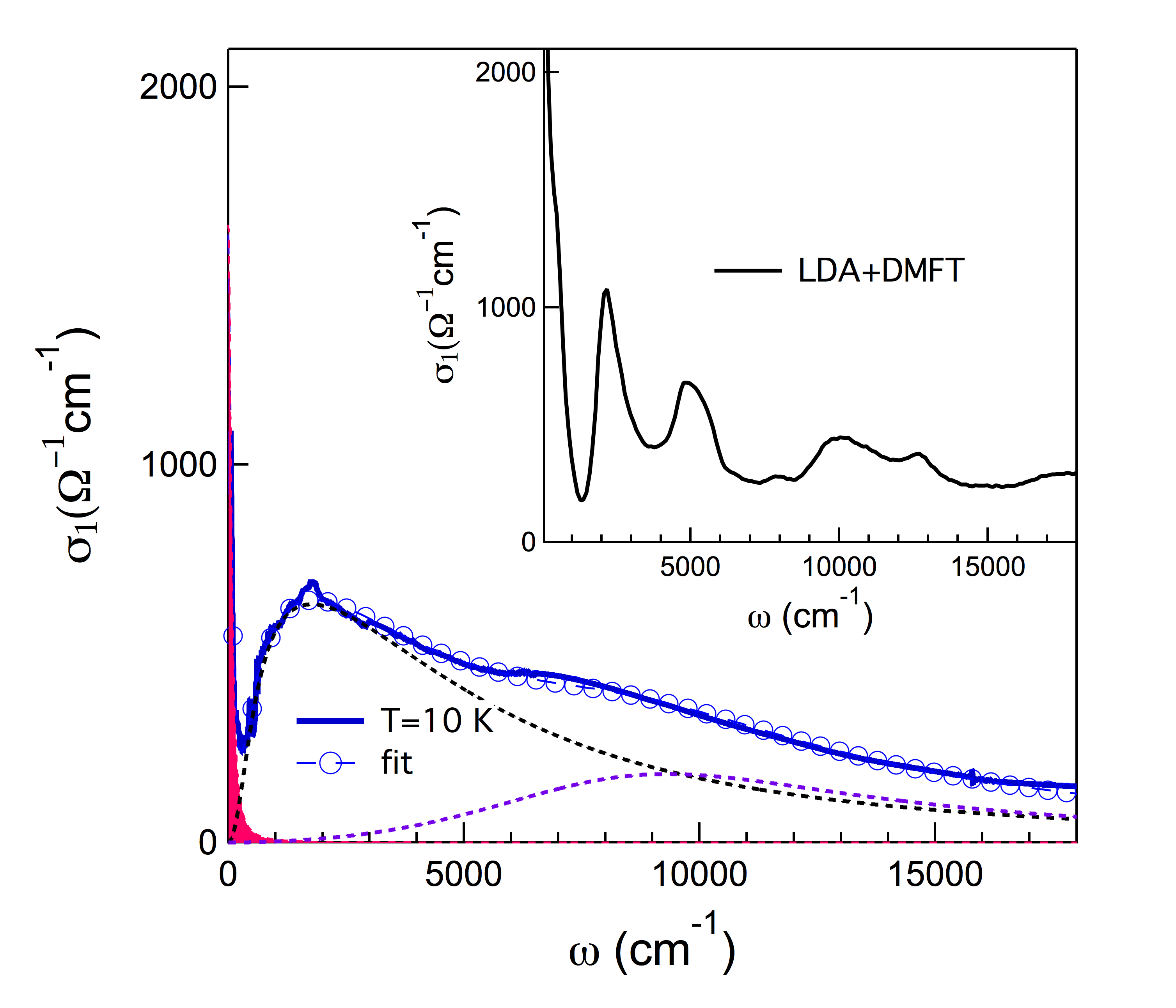}  
\end{center}
\caption{Drude-Lorentz fit of the optical conductivity data at 10 K. The fit components are a Drude peak (pink filled area), a mid-infrared absorption (dashed black line), and a high frequency Lorentzian (dashed purple line). The LDA+DMFT calculated curve using U=2.3, J$_H$=0.345 eV is shown in the inset.}
\label{Fig2}
\end{figure}

In Fig. \ref{Fig2} we report a Drude-Lorentz analysis of the optical conductivity at T=10 K. The complex optical conductivity is written in terms of a Drude contribution and Lorentzian oscillators as\cite{Dressel}:

\begin{equation}
\tilde{\sigma}(\omega)=\frac{\omega_P^2\tau}{4\pi(1-i\omega\tau)}+\frac{\omega}{4\pi i}\sum_j\frac{S_j^2}{\omega_j^2-\omega^2-i\omega\gamma_j}
\label{DrudeLorentz}
\end{equation}

In the Drude term $\omega_P$ is the plasma frequency, $\tau$ is the scattering time, while the Lorentz oscillators are peaked at finite frequencies $\omega_j$ with strength $S_j$ and width $\gamma_j$. We introduced a Lorentzian oscillator for the mid-infrared band and one for the higher-frequency component peaked around 9300 cm$^{-1}$. From the fitting parameters one can calculate the spectral weight ratio \cite{DeGiorgi-11}:

\begin{equation}
SW_{D}/SW_{D+MIR}=\frac{\omega_p^2}{\omega_p^2+S_{MIR}^2}
\label{Eq3}
\end{equation}

\noindent Where $SW_{D}$=$\omega_p^2$ is the spectral weight of the Drude component while $SW_{D+MIR}$ also includes the MIR contribution. This ratio provides an experimental estimate of the degree of electronic correlation of a material \cite{DeGiorgi-11,Baldassarre-12,LoVecchio-15}. If $SW_{D}/SW_{D+MIR}$ is small it means that electron-electron correlations are strongly renormalizing the coherent spectral weight, while for weakly correlated metals like gold and silver one obtains $SW_{D}/SW_{D+MIR}$$\to$1\cite{Qazilbash-09}. 
Computing the ratio from our measurements at 10 K we find $SW_{D}/SW_{D+MIR}$$\sim$0.03. This provides a clear and direct evidence that LiOsO$_3$ is a strongly correlated metal, where electron-electron repulsion hugely renormalizes the quasiparticle spectral weight in favor of the MIR incoherent part. 

\emph{Density Functional and Dynamical Mean Field Theory Calculations -}
In order to better characterize the strongly correlated metallic state of LiOsO$_3$, we compute the optical conductivity using the combination of DFT and DMFT. Comparing with the single particle spectral function, we can identify the electronic processes responsible for the various features in the optical absorption.
We start from Local Density Approximation (LDA \cite{LDA}) DFT calculations using the Vienna {\it Ab initio} Simulation Package (VASP) \cite{VASP} with the projector augmented wave (PAW) method \cite{PAW}.
To incorporate the role of electron-electron interactions in LiOsO$_3$ we build maximally localized Wannier orbitals out of the LDA bands \cite{wannier90} and we include local Coulomb interactions parameterized by the $U$ and the Hund's coupling $J_H$, which are accurately accounted for by DMFT. In the specific case of LiOsO$_3$, we construct Wannier orbitals for the $4d$ Os orbitals over the energy range spanned by the three $t_{2g}$ orbitals. 
As a DMFT solver, we use Exact Diagonalization (ED) \cite{Caffarel,Capone}, in which the impurity model is solved with a number of levels Ns$=$12, and diagonalized by a parallel Arnoldi algorithm \cite{ARPACK}.

\begin{figure}[!h]
\includegraphics[width=0.75\columnwidth,angle=-90]{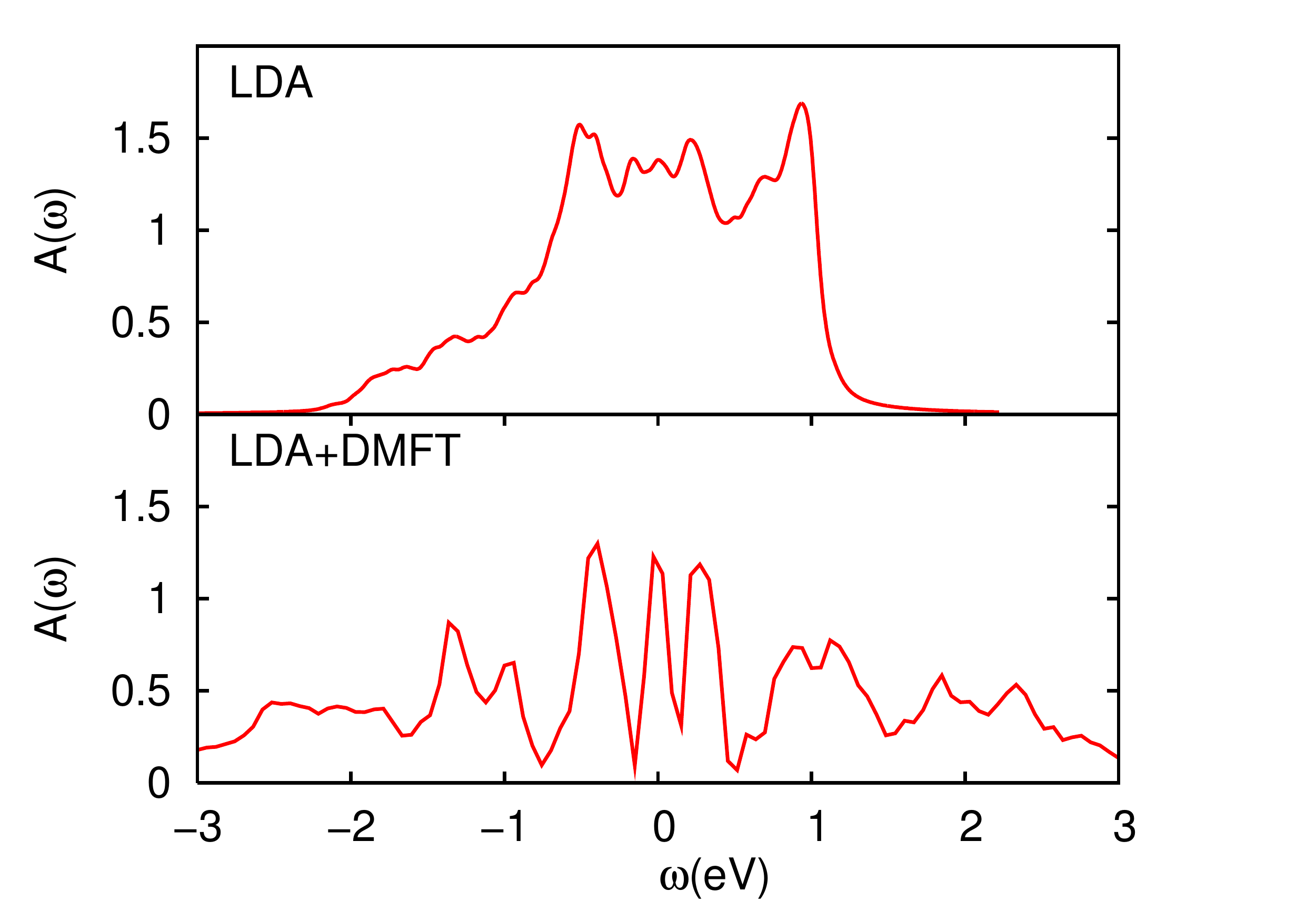}
\vspace{0.25cm}
\caption{Spectral density of LiOsO$_3$ calculated in LDA (a) and LDA+DMFT (b) with  U=2.3 eV and  J$_H$=0.345 eV.}
\label{Fig3}
\end{figure}

Using a description in terms of Os $t_{2g}$ orbitals, the electronic configuration of LiOsO$_3$ corresponds to a half-filled (three electrons per Os-site) $t_{2g}$ manifold of bands, a configuration which suffers from strong correlation effects even for moderate values of the screened Coulomb interaction, as anticipated in Ref. \onlinecite{Giovannetti-14}. Within LDA the bandwidth of $t_{2g}$ bands is close to 3.5 eV (see Fig. \ref{Fig3}), and the strong asymmetry due to the hybridization disfavors antiferromagnetism, which would turn the correlated metal into an insulator.

A naive estimate of the degree of correlation is simply to compare the non-interacting bandwidth with the on-site Coulomb interaction, which here can be estimated around 2 eV by comparison with similar compounds\cite{Giovannetti-15}. As these values might suggest an intermediate degree of correlation, it is important to take into account also the Hund's exchange coupling $J_H$, which, for half-filled multiorbital systems indeed favors the electronic localization, leading to a smaller critical value of $U$ for the Mott-Hubbard transition\cite{DeMediciJanus}.  As a result, for the physically relevant values of U and J$_H$ LiOsO$_3$ is indeed in a highly correlated regime, as discussed in Ref. \onlinecite{Giovannetti-15}. We now discuss how the strong correlations manifest themselves in the optical conductivity and we compare the theoretical results with experiments. 

Accurate values of U and J$_H$ are not known in LiOsO$_3$ and we are not aware of constrained RPA or alternative ab-initio estimates. Our prediction lies in the same range evaluated for the iridate compounds Sr$_2$IrO$_4$ and Ba$_2$IrO$_4$ \cite{irU}. In Fig. \ref{Fig3} we show the spectral density of states obtained within LDA and LDA+DMFT for the chosen interaction values, namely U$=$2.3 eV and $J_H$=0.345 eV (in the Kanamori notation for the exchange interactions). The effects of strong correlations clearly emerge by comparing the two plots. The relatively broad LDA band substantially shrinks due to electron-electron interactions, while spectral weight is shifted to high energy to form the so called Hubbard bands, centered roughly around $\pm U/2$.
The relative weight of the low energy excitations, measured by the quasiparticle weight $Z$ is close to 0.1, which is in remarkable agreement with the experimental estimate $SW_{D}/SW_{D+MIR}$.  The agreement between the experimental and theoretical conductivities is not limited to the Drude part as the overall distribution of spectral weight is reproduced by the calculations. Both in the experiment and in the theory we observe a broad mid-infrared peak  and a higher frequency feature. The former can easily be associated with optical excitations connecting the quasiparticle peak at the Fermi level with the lower and the upper Hubbard bands, while the latter is related to transitions from the lower to the upper Hubbard band, as one can easily realize comparing the optical spectrum with the single particle density of states. The agreement between the optical conductivity and the LDA+DMFT results clearly confirms a very important role of strong correlations in LiOsO$_3$\cite{Giovannetti-14}, which appears as a very bad metal on the verge of Mott-Hubbard localization.

As a matter of fact, the proximity of LiOsO$_3$ to a Mott-Hubbard transition and the consequent bad metallic behavior can be ascribed to the half-filled electronic configuration, where correlations are stronger due to the synergetic role of the Hubbard $U$ and the Hund's coupling. 
Such half-filled correlated electronic configuration is indeed common to Mott multiferroic materials \cite{SBMOGG,Nourafkan} in which however the ferroelectric state is realized in a strongly insulating system. While the presence of ferroelectric distortions in Mott insulators is not surprising, the existence of polar distortions in a metal is a remarkable result which still requires a complete understanding. 
However our characterization of the metallic state helps to reconcile the metallic state with the existence of polar distortions. 
The strongly correlated metallic state of LiOsO$_3$ can indeed be seen as an ``almost insulating" compound in which a very small metallic peak is present, while the high-energy spectrum is already that of a Mott insulator. The heavy quasiparticles at the Fermi level are expected to lead to a much less effective screening of the dipoles, cooperating with the Anderson-Blount mechanism\cite{Giovannetti-14,SEMOGG}. 

\emph{Conclusions -} We performed a combined theoretical and experimental study of the ferroelectic metal LiOsO$_3$. Infrared spectroscopy measurements are closely reproduced by calculations using the merger of Density Functional Theory and Dynamical Mean Field Theory. The optical spectra show a very small Drude weight associated with a strongly correlated character due to electron-electron correlation. The theoretical analysis provides a quasiparticle weight of 0.1, a much smaller value than in many other correlated metals.  Also the high energy features are interpreted in terms of excitations of a strongly correlated metal on the brink of a Mott-Hubbard metal-insulator transition. 

The first unambiguous ferroelectric metal emerges therefore as an extremely correlated compound, which could turn into a Mott insulator if the role of the interaction was only slightly stronger\cite{Giovannetti-14,Giovannetti-15}. We argue that the proximity to such a metal-insulator transition is indeed a key property which favors the existence of polar distortion in a metallic state, suggesting that the poor metal close to Mott localization is not able to fully screen the electric dipoles.

GG thanks D. Puggioni for useful discussions.
GG and MC acknowledge financial support by European Research Council
under FP7/ERC Starting Independent Research Grant ``SUPERBAD" (Grant Agreement n. 240524). KY thanks a Grant-in-Aid for Scientific Research (25289233 and 15K14133) from JSPS. The research leading to these results has received funding from the European Community’s Seventh Framework Programme (FP7/2007-2013) under grant agreement nº 312284.


\begin{thebibliography}{}
\bibitem{note70} For 200 K and 300 K the lower end of our frequency range was 70 cm$^{-1}$.
\end{thebibliography}
\end{document}